\documentclass[iop]{emulateapj}
\usepackage{amsmath}
\usepackage{graphicx}
\usepackage{natbib}

\def\gtwid{\mathrel{\raise.3ex\hbox{$>$\kern-.75em\lower1ex\hbox{$\sim$}}}}
\def\ltwid{\mathrel{\raise.3ex\hbox{$<$\kern-.75em\lower1ex\hbox{$\sim$}}}}
\def\\{\hfil\break}

\def\lesssim{\mathrel{\hbox{\rlap{\hbox{\lower4pt\hbox{$\sim$}}}\hbox{$<$}}}}
\def\gtrsim{\mathrel{\hbox{\rlap{\hbox{\lower4pt\hbox{$\sim$}}}\hbox{$>$}}}}

\newcommand{\mamo}[1]{\mbox{$#1$}}
\newcommand{\unit}[1]{\ifmmode \:\mbox{\rm #1}\else \mbox{#1}\fi}
\newcommand{\sbr}[1]{_{\rm #1}}

\newcommand{\mone}{\mamo{^{-1}}}

\newcommand{\kms}{\unit{km~s\mone}}

\newcommand{\chisq}{\mamo{\chi^2}}

\newcommand{\jcap}{JCAP}

\newcommand{\Om}{\mamo{\Omega\sbr{m}}}
\newcommand{\Omzero}{\mamo{\Omega_{\mathrm{m},0}}}

\newcommand{\LCDM}{$\Lambda$CDM}

\slugcomment{ApJ Letters, in press.}

\begin{document}

\title{The growth rate of cosmic structure from peculiar velocities at low and high redshifts}
\author{Michael J. Hudson\altaffilmark{1} and Stephen J. Turnbull}
\affil{Dept. of Physics and Astronomy, University of Waterloo, Waterloo, Ontario, Canada.}
\email{mjhudson@uwaterloo.ca}
\altaffiltext{1}{Perimeter Institute for Theoretical Physics, Waterloo, Ontario, Canada.}
\shorttitle{Growth rate of cosmic structure from peculiar velocities}
\shortauthors{Hudson \& Turnbull}

\begin{abstract}
Peculiar velocities are an important probe of the growth rate of mass density fluctuations in the Universe. Most previous studies have focussed exclusively on measuring peculiar velocities at intermediate ($0.2 < z < 1$) redshifts using statistical redshift-space distortions. Here we emphasize the power of peculiar velocities obtained directly from distance measurements at low redshift ($z \lesssim 0.05$), and show that these data break the usual degeneracies in the \Omzero -- $\sigma_{8,0}$ parameter space. Using only peculiar velocity data, we find $\Omzero = 0.259\pm0.045$ and $\sigma_{8,0} = 0.748\pm0.035$.  Fixing the amplitude of fluctuations at very high redshift using observations of the Cosmic Microwave Background (CMB), the same data can be used to constrain the growth index $\gamma$, with the strongest constraints coming from peculiar velocity measurements in the nearby Universe. We find  $\gamma = 0.619\pm 0.054$, consistent with \LCDM{}. Current peculiar velocity data already strongly constrain modified gravity models, and will be a powerful test as data accumulate.
\end{abstract}

\maketitle

\label{firstpage}

\section{Introduction}

In the standard cosmological model, the Universe is dominated by cold dark matter combined with a cosmological constant or dark energy (\LCDM).
While the existence of dark matter is supported by dynamical tests, gravitational lensing and fluctuations in the Cosmic Microwave Background (CMB), the evidence for a cosmological constant is primarily through geometric  constraints on the redshift-distance relation (standard candles, standard rulers) and hence on the expansion history of the Universe \cite[for example][hereafter WMAP7+BAO+$H_{0}$]{KomSmiDun11}. Modified gravity theories, however, can mimic the expansion history of the \LCDM{} model.  \cite{Lin05} has emphasized that it is essential to measure the growth of structure as a function of cosmic time as this allows one to break this degeneracy.  He also shows that for many models, the logarithmic derivative of the growth of structure can be parametrized as
\begin{equation}
f (z)\equiv \frac{d \ln g}{d \ln a} = \Om(z)^{\gamma}
\end{equation}
where $z$ is the redshift, $g$ is the linear perturbation growth factor, $a = 1/(1+z)$ is the expansion factor and $\gamma$ is 0.55 for \LCDM{}  \citep{WanSte98}. In contrast, for example, $\gamma = 0.68$ in the \citet[][hereafter DGP]{DvaGabPor00} braneworld modified gravity model \citep{LinCah07}.

There are several ways to measure the amplitude of the dark matter power spectrum at redshifts lower than that of the CMB, including cosmic shear from weak gravitational lensing and the abundance of rich clusters.
Another promising way to probe the growth \emph{rate} of structure is via peculiar velocities \citep{GuzPieMen08, KosBha09}. Peculiar velocities are directly proportional to the derivative of the growth factor, i.e.\ proportional to $f$.

There are two ways to measure peculiar velocities. The first method is statistical: given a galaxy redshift survey, the distortion of the power spectrum or correlation function in redshift space depends on $\beta = f/b$, where $b$ is a galaxy bias parameter \citep{Kai87}.  On large scales, we may assume that linear biasing holds, i.e. $b = \sigma_{8,g}/\sigma_{8}$, where $\sigma_{8}$ is the root-mean-square density contrast within an 8 Mpc/$h$ sphere, $h$ is the Hubble parameter in units of 100 km/s/Mpc, and the subscript ``g'' indicates the fluctuations in the galaxy density, whereas no subscript indicates fluctuations in the mass density contrast.  Galaxy redshift surveys also allow one to measure $\sigma_{8,g}$ directly, so one can combine the observables to obtain the combination $f\sigma_{8} = \beta \sigma_{8,g}$.  By combining redshift-space distortion (RSD) measurements of $f\sigma_{8}$ at different redshifts, one can study the growth of linear structures over a range of redshifts \citep{GuzPieMen08, SonPer09, BlaBroCol11, SamPerRac12}.

A second method is to measure peculiar velocities directly, by measuring distances to individual galaxies (via standard candles or standard rulers), and comparing these distances to their redshifts. We refer to this method as ``measured distance" (MD)%
. The measured peculiar velocities can then be compared with the peculiar velocities predicted from a galaxy density field, $\delta_g(\mathbf{r})$,  derived from an independent redshift survey, under the assumption that fluctuations in the mass are linearly related to those in the galaxy density. Specifically, we have 
 \citep{PPC}
\begin{equation}
\mathbf{v}(\mathbf{r}) = \frac{H_0}{4 \pi}\frac{f}{b} \int_{0}^{\infty} d^3\mathbf{r}'\delta_g(\mathbf{r}') \frac{\mathbf{r}'-\mathbf{r}}{\mid \mathbf{r}'-\mathbf{r}\mid^{3}}\,.
 \label{eq:vpred}
\end{equation} 
This comparison of the two measured quantities $\mathbf{v}(\mathbf{r})$ and $\delta_g(\mathbf{r})$ yields a measurement of the degenerate combination  $\beta = f/b$, which can be converted into the combination $f \sigma_{8}$ in the same way as for RSDs. Note that because this is a velocity-density cross-correlation, unlike RSDs, it is not affected by cosmic variance.  Instead,  the noise arises from the precision of the MDs themselves.

The two peculiar velocity probes are complementary: RSDs require large volumes, driving one to surveys at higher redshifts.  MDs have errors which are a constant fraction of distance. Hence the error in peculiar velocity in units of \kms{} increases linearly with distance and so MD surveys are necessarily restricted to low redshifts. However, as we will show it is the lowest redshift data that have the most ``lever arm'' for constraining the cosmological parameters considered here. The important point is that by combining high and low redshift measurements of $f (z) \sigma_{8}(z)$, we can break many degeneracies in the cosmological parameters.

An outline of this Letter is as follows. In Section 2, we present the data used in our analysis. Section 3 presents fits in the $\Omzero$-$\sigma_{8,0}$ plane for the \LCDM{} class of models.  In Section 4, we allow $\gamma$ to vary, but use CMB data to constrain the amplitude of $\sigma_{8}(z\sbr{CMB})$. We discuss future prospects in Section 5 and summarize in Section 6.

\section{Data}

We will combine measurements of $f\sigma_{8}$ from the two distinct methods discussed above.  The majority of the data are from RSD measurements, many of which were previously compiled in \cite{BlaBroCol11}, but adding recent results from BOSS \citep{ReiSamWhi12} and 6dFGS \citep{BeuBlaCol12}. These are summarized  in Table 1.

\begin{deluxetable}{lcccl}
\tablecaption{Measurements of $f\sigma_8$ from the literature, and derived $\gamma$}
\tablehead{\colhead{Label} & \colhead{$z$} & \colhead{$f\sigma_8$} & \colhead{$\gamma$} & \colhead{Ref}}
\startdata
THF & 0.02 & $0.398\pm0.065$ & $0.56_{-0.09}^{+0.11}$ & 1 \\
DNM & 0.02 & $0.314\pm0.048$ & $0.71_{-0.09}^{+0.10}$ & 2 \\
6dF & 0.07 & $0.423\pm0.055$ & $0.54_{-0.08}^{+0.09}$ & 3 \\
2dF & 0.17 & $0.510\pm0.060$ & $0.43_{-0.08}^{+0.10}$ & 4 \\
LRG1 & 0.25 & $0.351\pm0.058$ & $0.77_{-0.14}^{+0.16}$ & 5 \\
LRG2 & 0.37 & $0.460\pm0.038$ & $0.55_{-0.08}^{+0.09}$ & 5 \\
WZ1 & 0.22 & $0.390\pm0.078$ & $0.67_{-0.15}^{+0.19}$ & 6 \\
WZ2 & 0.41 & $0.428\pm0.044$ & $0.64_{-0.11}^{+0.12}$ & 6 \\
WZ3 & 0.6 & $0.403\pm0.036$ & $0.76_{-0.13}^{+0.14}$ & 6 \\
WZ4 & 0.78 & $0.493\pm0.065$ & $0.38_{-0.24}^{+0.28}$ & 6 \\
BOSS & 0.57 & $0.415\pm0.034$ & $0.71_{-0.11}^{+0.12}$ & 7 \\
VVDS & 0.77 & $0.490\pm0.180$ & $0.40_{-0.59}^{+0.89}$ & 8
\enddata
\tablecomments{The \LCDM{} distance-redshift relations are assumed when calculating the AP distortion, but the growth factor $f\sigma_8$ is free. Derived $\gamma$ also uses WMAP measurement of fluctuations at $z\sbr{CMB}$}  
\tablerefs{{(1) \cite{TurHudFel12}; (2) \cite{DavNusMas11}; (3) \cite{BeuBlaCol12}; (4) \cite{SonPer09}; (5) \cite{SamPerRac12}; (6) \cite{BlaGlaDav11}; (7) \cite{ReiSamWhi12}; (8) \cite{GuzPieMen08,SonPer09}}}
\end{deluxetable}

As discussed above, it is also possible to measure $f\sigma_8$ from the comparison of peculiar velocity and density fields, where peculiar velocities are obtained directly from MDs. Most of the results from this method (up to 2005) were summarized in Table 3 of \citet[hereafter PH]{PikHud05}. They showed that results from estimates of $f\sigma_8$ were consistent\footnote{After correcting for their assumed $\gamma=0.6$ and making an additional correction from the non-linear $\sigma_8$ using \cite{JusFelFry10} and assuming $\Omzero =  0.27$. 
} with $0.41\pm0.03$. However, since both the peculiar velocity and redshift survey datasets overlap somewhat, the measurements are not independent and it is therefore difficult to assess the uncertainties.

Its possible to obtain a value of $f\sigma_8$ using non-overlapping datasets: by comparing peculiar velocities from the ``Composite'' sample of \cite{WatFelHud09} to predictions derived from the IRAS Point Source Catalog Redshift (PSCz) survey galaxy density field \citep{SauSutMad00}, we find $\beta\sbr{I}= 0.49\pm0.04$ after marginalizing over residual bulk flows.  This yields $f \sigma_8 = 0.37\pm0.04$, in good agreement with the PH mean value quoted above.  These uncertainties may underestimate possible systematics, however, since only one density field (PSCz) is used and, furthermore, the same comparison method is used for all peculiar velocity samples.

Therefore, in order to make a fair (albeit conservative) estimate of the uncertainties in the methods, in this Letter we focus on only two recent measurements of $f \sigma_{8}$ from MD surveys. The first of these is from \citet[hereafter THF]{TurHudFel12}, who compiled the `First Amendment' (hereafter A1) set of 245 peculiar velocities from Type Ia supernovae (SNe) with $z < 0.067$.  The A1 peculiar velocities were compared to predictions from the IRAS PSCz galaxy density field, yielding $f\sigma\sbr{8,lin} = 0.40 \pm 0.07$ at a characteristic depth of $z = 0.02$.  We note that the uncertainties on $f \sigma_{8}$ are marginalized over possible residual bulk flows.  The results are also insensitive to details of, for example, corrections for extinction in the SN host\footnote{Changing to $R_V = 3.1$ instead of the default $R_V = 1.7$ changes $f \sigma_{8}$  by only 0.05. \cite{KesBecCin09} fit $R_{V} = 2.18\pm0.14\sbr{stat}\pm0.48\sbr{syst}$, so the difference between $R_{V} = 1.7$ and $R_{V} = 3.1$ is a factor 2.8 times the systematic error.  Hence the 1-$\sigma$ $R_{V}$ systematic effect on $f\sigma_{8}$ ($\sim 0.02$) is much smaller than the random errors ($0.07$)}. Further details about the uncertainties and light curve fitters used for the A1 SNe can be found in THF.  

The second low-$z$ MD result is from \citet[hereafter DNM]{DavNusMas11}, who analyzed 2830 Tully-Fisher peculiar velocities at $z < 0.033$ and compared these to the predictions from the galaxy density field derived from the Two Micron All-Sky Survey Redshift Survey \citep{HucMacMas12}. The analysis differs from THF: rather than the simple point-by-point comparison of predicted and observed peculiar velocities, DNM applied a spherical harmonic decomposition to both fields and found $f\sigma_{8} = 0.31\pm0.05$.  We note that a straight average of the THF and DNM results yields $f\sigma_{8}=0.36\pm0.04$ in excellent agreement with the Composite vs. PSCZ result found above. 

\begin{figure}
  \includegraphics[width=\columnwidth]{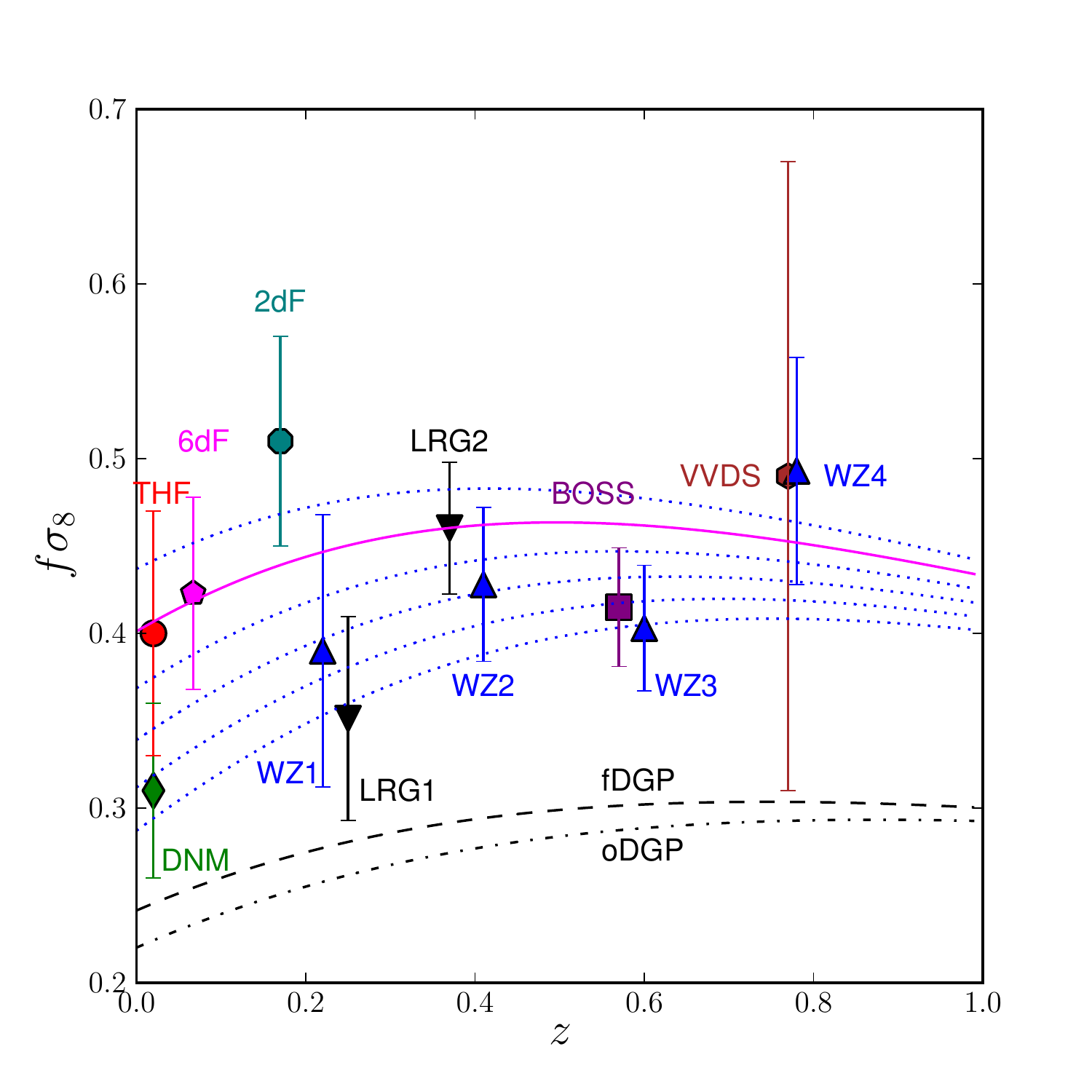}   
  \caption{Growth parameter $f\sigma_8(z)$ as a function of $z$. The data and errorbars are labelled as in Table 1. The \LCDM{} model with WMAP7+BAO+$H_{0}$ parameters $\Omzero = 0.275$, $\sigma_{8,0} = 0.816$  is shown by the solid magenta curve.  Note that the high-redshift RSD points assume the \LCDM{} redshift-distance relation to correct for the AP effect and hence the appropriate value of $f\sigma_8(z)$.
  The dotted curves are normalized to have the correct $\sigma_{8}$ at $z\sbr{CMB}$ but have different possible $\gamma$ values starting at 0.50 at the top and increasing in steps of 0.05 towards the bottom. The dashed and dash-dot curves show predictions of flat and open DGP models, respectively, normalized to the amplitude of the fluctuations in the CMB (see text for details). \label{fig:Fsig8}}
\end{figure}

We stress the that these two distance measurement results are completely independent: they use different peculiar velocity samples, different density fields and different reconstruction and comparison methods.  The two measurements are consistent with each other: the difference in $f\sigma_{8}$ is $0.09\pm 0.08$.   The RSD and MD measurements are shown in Figure \ref{fig:Fsig8}.

\section{Determination of $\Omzero$ and $\sigma_{8,0}$}

In the \LCDM{} model, $\gamma = 0.55$. The model allows us to predict the value of $f(z) = \Om(z)^{\gamma}$ and $\sigma_8(z)$ at any redshift, assuming their values at redshift $z=0$ (denoted by subscript ``0'').  

For the fits, we use a simple $\chi$ squared statistic with the following form:
\begin{equation}
 \chisq = \sum_{i} \frac{[f\sigma_{8}(\mathrm{meas})_i-f\sigma_8(\mathrm{pred})_i]^2}{\sigma_{i}^{2}}
 \label{eq:Chi2}
\end{equation}
where the first term is the measured value and the second term is the model, and $\sigma_{i}^{2}$ is the uncertainty for each measured value.  For RSD measurements at $z \gtrsim 0.1$, $f\sigma_8$ is degenerate with the redshift-distance relation, due to the \citet[hereafter AP]{AlcPac79} effect.  If we assume a flat Universe, then the redshift-distance relation, and hence the AP distortion are all fixed by our choice of $\Omzero$. With the AP distortion fixed, the appropriate $f\sigma_{8}$ values can then be calculated, but only for those surveys that have published the covariance matrix between the AP effect and $f\sigma_8$: WZ and BOSS.  The peculiar velocity measurements (THF and DNM) and the low-$z$ 6dF RSD measurement are negligibly affected by the AP distortion.

\begin{figure}
  \includegraphics[width=\columnwidth]{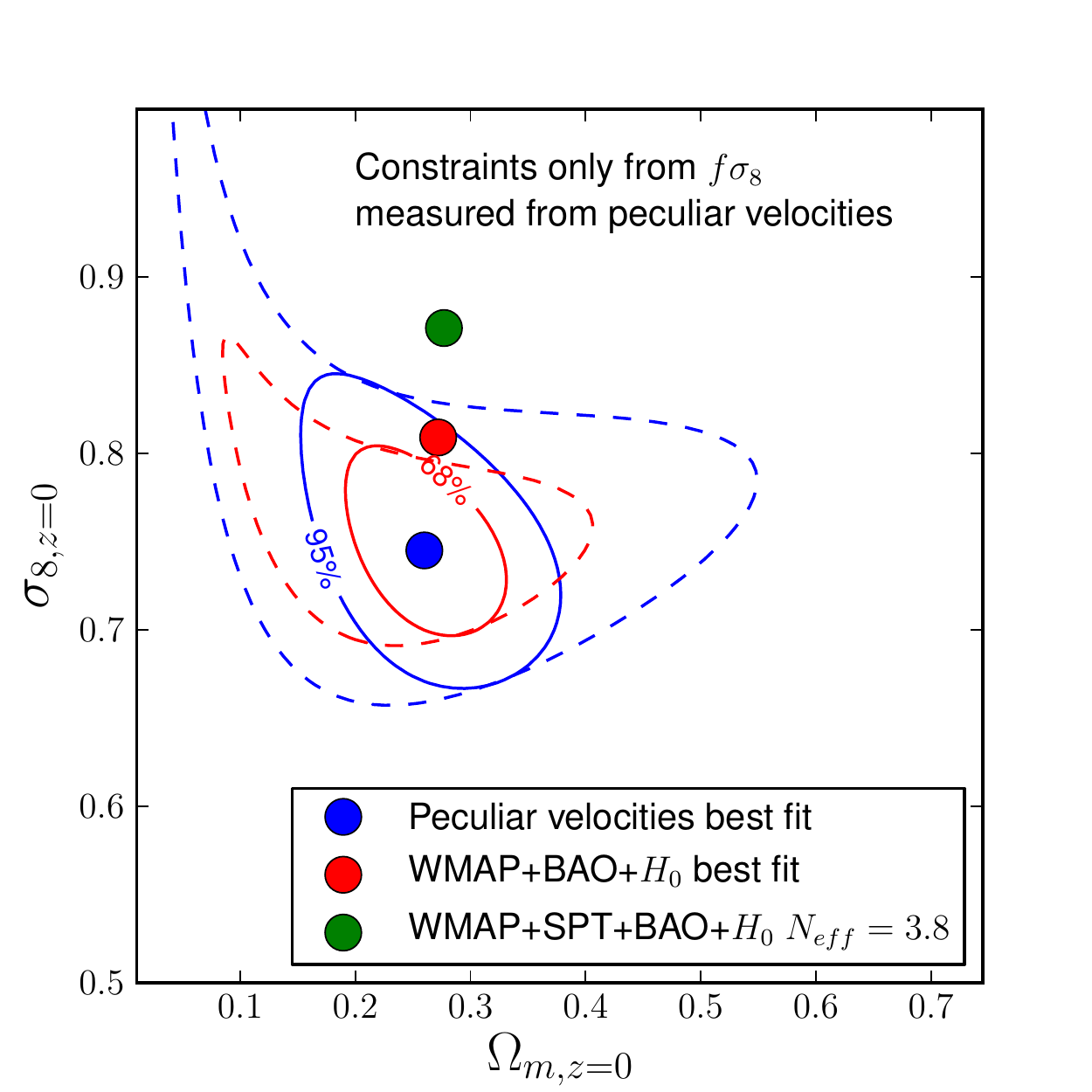}
  \caption{$\Omzero$ versus $\sigma_{8,0}$. Dashed contours show the 68, 95 and 99 per cent confidence intervals using only WZ and BOSS  redshift-space distortion measurements.  The solid contours show the same constraints for the above data plus the peculiar velocity measurements (THF and DNM) and the low-$z$ 6dF RSD measurement (which is unaffected by the AP effect). The red circle shows the WMAP7+BAO+$H_{0}$ best fit parameters for comparison, the blue circle shows our best fit. The green circle shows the WMAP+SPT+BAO+$H_0$ best fit for a model with 3.8 effective relativistic species \citep{KeiReiAir11}, which is disfavored at $>99$ \% confidence level.
\label{fig:omsig8}}
\end{figure}

Figure \ref{fig:omsig8} shows the results of the fit in the $\Omzero$ versus $\sigma_{8,0}$ parameter plane with $\gamma = 0.55$. While the WZ+BOSS-only fit is quite degenerate in the $\Omzero-\sigma_{8,0}$ plane, these degeneracies are broken when low-$z$  data are included in the fit. Assuming a fixed $\gamma = 0.55$, the best fit is $\Omzero = 0.259\pm0.045$ and $\sigma_{8,0} = 0.748\pm0.035$ but the WMAP7+BAO+$H_{0}$ value is consistent with these results. Some \LCDM{} variants, such as those with  a non-standard effective number of relativistic species, are disfavored. The figure shows one such example ($N\sbr{eff}=3.8$), from \citet{KeiReiAir11}. We stress that these results depend only on growth measurements and hence are independent of other determinations (CMB, SNe, BAO etc.)

\section{Constraints on $\gamma$}

As discussed above, it is also interesting to measure the growth rate index $\gamma$. However, once $\gamma$ is included as a third parameter, the fits become very degenerate. These degeneracies can be broken using CMB data, for example from WMAP7+BAO+$H_0$. Let us assume that the Friedman equation (and hence the expansion history) is the same as for the \LCDM{} model, but treat $\gamma$ as a phenomenological growth parameter which is allowed to differ from 0.55. At high redshifts ($z \sim 1000$), $\Om = 1$ to high accuracy and so the growth of perturbations at early epochs is independent of $\gamma$.  Therefore, we can use WMAP7+BAO+$H_0$  parameters\footnote{While the primary fluctuations in the CMB are independent of $\gamma$, the secondary Integrated Sachs-Wolfe (ISW) effect is not. The ISW effect affects the CMB anisotropy spectrum on very large angular scales. We neglect this effect here.} \Omzero\ and  $\sigma_{8,0}$ to fix the amplitude of fluctuations at high redshift, as well as fixing the Friedman equation and expansion history. Note that the quoted WMAP7  $\sigma_{8,0}$  is extrapolated to $z=0$ assuming \LCDM. To allow for other values of $\gamma$ we use \LCDM{} to calculate $\sigma_8(z\sbr{CMB})$, and then extrapolate the $z\sbr{CMB}$ predictions forward at later times, using different values of $\gamma$ \citep[following][Section 4.5]{SamPerRac12}. We then use the WMAP7+BAO+$H_{0}$ Monte Carlo Markov chains to marginalize over \Omzero\ and $\sigma_{8,0}$. The resulting fits are shown in Fig. \ref{fig:Fsig8}.

\begin{deluxetable}{lcccl}
\tablecaption{Measurements of $\gamma$ from combinations of the data}
\tablehead{\colhead{Sample} & \colhead{$\gamma$} & \colhead{$\sigma_{\mathrm{CMB}}$} & \colhead{$\sigma_{\mathrm{tot}}$}}
\startdata
WZ & $0.666^{+0.077}_{-0.073}$ & 0.053 & 0.092 \\
LRG & $0.625^{+0.072}_{-0.077}$ & 0.046 & 0.088 \\
All RSD & $0.607^{+0.038}_{-0.040}$ & 0.046 & 0.060 \\
THF+DNM & $0.653^{+0.073}_{-0.064}$ & 0.035 & 0.077 \\
\bf All & $\mathbf{0.619^{+0.033}_{-0.035}}$ &  \bf 0.042 & \bf 0.054\\
\enddata
\end{deluxetable}

\begin{figure}
  \includegraphics[width=\columnwidth]{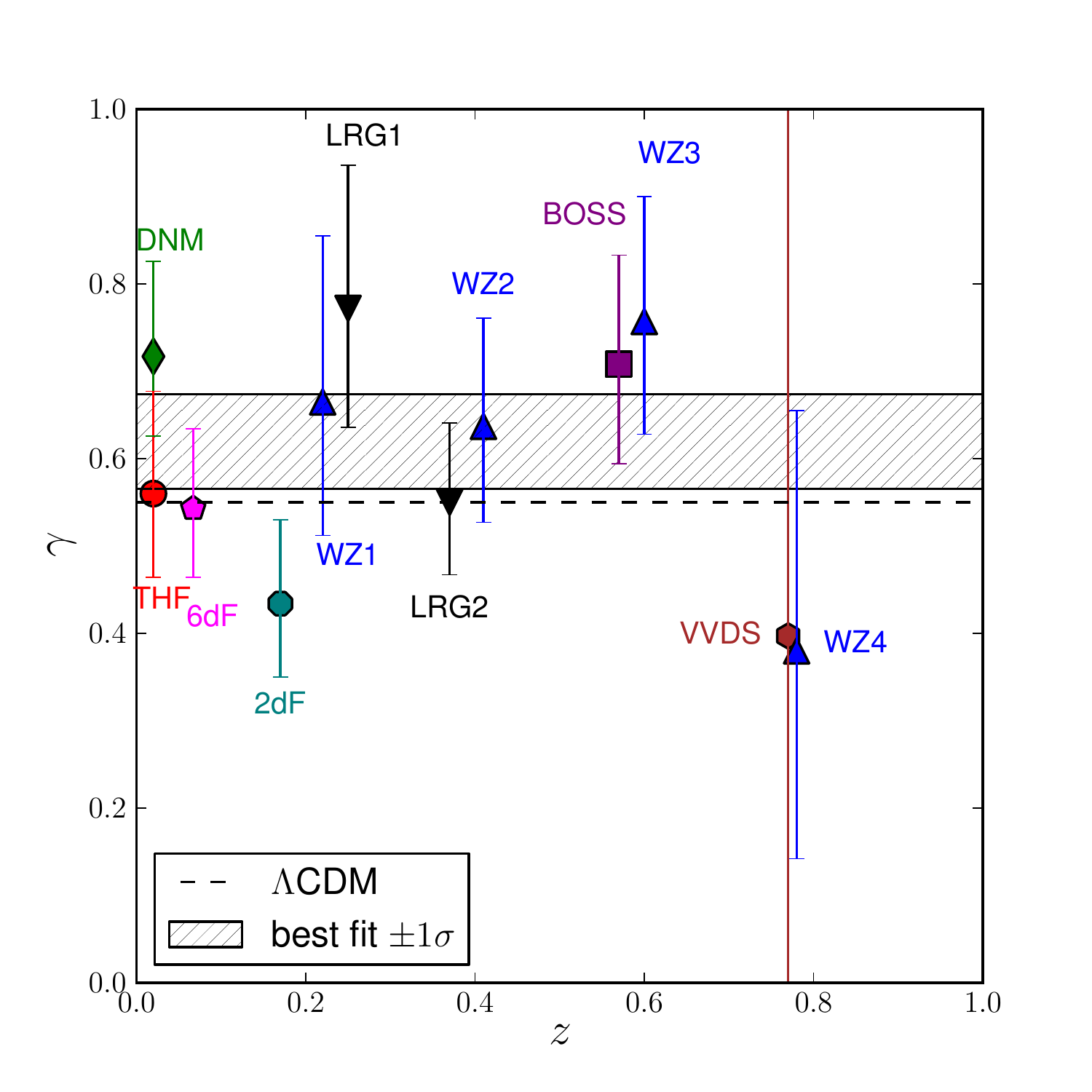}
\caption{Values of $\gamma$ derived for each survey individually, assuming the WMAP7+BAO+$H_{0}$ central parameters and extrapolating the WMAP7 $\sigma_{8,0}$ to $z\sbr{CMB}$.  Error bars on individual measurements reflect only the uncertainties in the $f\sigma_{8}$ measurements and not the (correlated) uncertainties in the CMB-derived parameters. Symbols are as in Figure \ref{fig:Fsig8}.  The horizontal dashed line indicates the \LCDM{} expectation. The best fit to the data and $\pm1 \sigma$ range, including uncertainties in CMB-derived parameters is shown by the hatched box.} \label{fig:Gamma}
\end{figure}

Table 2 gives the derived $\gamma$ measurements for different combinations of the $f\sigma_{8}$ measurements. Also listed are the uncertainties in $\gamma$ arising from $f\sigma_{8}$, from the CMB-determinations of $\Omega\sbr{m,0}$ and $\sigma_{8,0}$, and the total error. Note that the errors from WMAP+BAO+$H_0$ are not independent (between one $f\sigma_{8}$ measurement and another), are weakly dependent on $z$, and dominate the total error budget when all data are combined. Figure \ref{fig:Gamma} shows the derived $\gamma$ for all $f\sigma_{8}$ measurements assuming the WMAP7+BAO+$H_{0}$ parameters, and fixing $\sigma_{8}(z\sbr{CMB})$. The results from the MD surveys are consistent with those from all of the RSD measurements.  The hatched region shows the best fit to all data\footnote{If we replace LRG1, LRG2 and BOSS with the ``free-growth'' fits from \cite{TojPerBri12}, we obtain a $\gamma$ only 0.013 lower.} combined: $0.619\pm0.054$, consistent with \LCDM.

Although for purposes of illustration we have focussed on a constant scale-independent $\gamma$, peculiar velocity data also allow one to test more complicated modified gravity scenarios \citep[see e.g.\ the review by][]{CliFerPad12}. One example is the DGP braneworld model. Fig. 1 shows the predicted values of $f\sigma_{8}$ for a self-accelerating flat DGP model, with parameters chosen to match the expansion history and CMB constraints \citep{LomHuFan09}, and with a perturbation amplitude at early times chosen to match the CMB.  Although the expansion history of the DGP model is similar to \LCDM{}, the Friedmann equation, and hence $\Omega_{m}(z)$, differs. Furthermore, its $\gamma = 0.68$ also differs considerably from the \LCDM{} value. These effects lead to lower growth and much lower values of  $f\sigma_{8}$ at $z <1$. This model, already disfavored at the $5\sigma$ level from other data \citep{LomHuFan09}, is excluded at the $\sim 8 \sigma$ level using only the $f\sigma_{8}$ measurements discussed here.  An open DGP model, which is a better fit to the redshift-distance relation, is excluded at the $\sim 10 \sigma$ level by growth measurements alone. This suggests that alternative modified gravity models will be strongly constrained by the requirement of simultaneously matching the \LCDM{} expansion history and the \LCDM{} growth history.

\section{Discussion}

The prospects for improving the measurements of $\gamma$ are excellent. At present, the error budget is \emph{dominated} by the WMAP7+BAO+$H_{0}$ estimates of the parameters at high redshift.  {\sc Planck} will reduce the CMB uncertainties so that these become subdominant. 

Peculiar velocity measurements will continue to improve, leading to a reduction in the observational errors in $f\sigma_8$. The BOSS measurement will improve with further data releases. RSD measurements are also being made at higher redshifts \citep{BieHilSha12}. 
However the statistical power \emph{all} of the RSD measurements combined (which are based on 800,000 redshifts) is similar to that of MD (based only on $\sim 3000$ peculiar velocity measurements), the subsamples having uncertainties of 0.060 and 0.077 respectively in $\gamma$ (Table 2). It is therefore clearly of great interest to improve the statistics of MDs. At low redshift, supernovae will continue to accumulate. We can also look forward to Fundamental Plane peculiar velocities from 6dFGS \citep{SprMagPro12}, and later an order of magnitude increase in the number of Tully-Fisher peculiar velocities from WALLABY\footnote{http://www.atnf.csiro.au/research/WALLABY/}.  Finally, the kinetic Sunyaev-Zel'dovich effect will be used to measure the velocity field of clusters at intermediate redshift \citep{HanAddAub12}. The redshift survey data used to construct the density field and hence the predicted peculiar velocities is also improving \citep{LavHud11}. It remains to better understand systematics, by comparing measurements using the same sets of peculiar velocity and density data, but different methods.

\section{Conclusion}
 
We have shown that by combining measurements of $f\sigma_8(z)$ at different redshifts, and in particular by including results at $z \sim 0$ from MD surveys, we can break the degeneracy between $\Omzero$ and $\sigma_{8,0}$ and obtain $\Omzero = 0.259\pm0.045$ and $\sigma_{8,0} = 0.748\pm0.035$, consistent with independent determinations from WMAP7+BAO+$H_{0}$.

We can also constrain the growth index $\gamma$ by comparing measurements of $f\sigma_8(z)$ at low $z$, after fixing their values at $z\sbr{CMB}$. The strongest leverage on $\gamma$ arises from peculiar velocity measurements at the lowest redshifts. By including these measurements, we derive $\gamma = 0.619\pm0.054$, consistent with \LCDM{}.  The {\sc Planck} mission plus upcoming peculiar velocity and redshift surveys will tighten these constraints further.

\section*{Acknowledgements}

We thank Gigi Guzzo, Florian Beutler and David Parkinson for useful comments. MJH and SJT acknowledge the financial support of NSERC and OGS respectively.

\bibliographystyle{apj}

\label{lastpage}

\end{document}